\documentclass[aps,amsmath,preprint,epsf,superscriptaddress,nofootinbib]{revtex4-1}
\pdfoutput=1
\usepackage{graphicx}
\usepackage{bm}
\usepackage{color}
\usepackage{amsmath,amssymb,amsfonts}

\usepackage{hyperref}
\hypersetup{
	colorlinks = true,
	linkcolor  = red, 
	linktocpage=true,
	citecolor={blue}
}

\usepackage{setspace}
\usepackage[samesize]{cancel}
\usepackage{tikz}
\usepackage{tikz-feynhand}
\usetikzlibrary{shapes}

\usepackage[utf8]{inputenc}
\usepackage{graphicx}
\usepackage{physics}
\usepackage{slashed}
\usepackage{relsize}
\usepackage[all]{xy}
\usepackage{verbatim}
\newcommand{\rr}[1]{{\color{red}#1}}
\newcommand{\ab}{{\alpha\beta}}
\newcommand{\cd}{{\gamma\delta}}

\makeatletter
\def\@ssect@ltx#1#2#3#4#5#6[#7]#8{%
  \def\H@svsec{\phantomsection}%
  \@tempskipa #5\relax
  \@ifdim{\@tempskipa>\z@}{%
    \begingroup
      \interlinepenalty \@M
      #6{%
       \@ifundefined{@hangfroms@#1}{\@hang@froms}{\csname @hangfroms@#1\endcsname}%
       {\hskip#3\relax\H@svsec}{#8}%
      }%
      \@@par
    \endgroup
    \@ifundefined{#1smark}{\@gobble}{\csname #1smark\endcsname}{#7}%
  }{%
    \def\@svsechd{%
      #6{%
       \@ifundefined{@runin@tos@#1}{\@runin@tos}{\csname @runin@tos@#1\endcsname}%
       {\hskip#3\relax\H@svsec}{#8}%
      }%
      \@ifundefined{#1smark}{\@gobble}{\csname #1smark\endcsname}{#7}%
      \addcontentsline{toc}{#1}{\protect\numberline{}#8}%
    }%
  }%
  \@xsect{#5}%
}%
\makeatother

\begin{document}
\baselineskip 0.7cm

\preprint{ULB-TH/23-02}
\bigskip

\title{A forgotten fermion: the hypercharge $-3/2$ doublet, its phenomenology and connections to dark matter}
\author{Rupert Coy}
\email[e-mail: ]{rupert.coy@ulb.be}
\affiliation{Service de Physique Théorique, Université Libre de Bruxelles,\\
Boulevard du Triomphe, CP225, 1050 Brussels, Belgium}

\begin{abstract}
\vspace{0.1cm}
A weak-doublet with hypercharge $-3/2$ is one of only a handful of fermions which has a renormalisable interaction with Standard Model fields. 
This should make it worthy of attention, but it has thus far received little consideration in the literature. 
In this paper, we perform a thorough investigation of the phenomenology which results from the introduction of this field, $F$. 
After expressing the model in terms of its effective field theory at dimension-6, we compute a range of electroweak and leptonic observables, the most stringent of which probe up to $M_F \sim 300$ TeV. 
The simplicity of this scenario makes it very predictive and allows us to correlate the different processes. 
We then study how this new fermion can connect the SM to various simple but distinct dark sectors. 
Some of the most minimal cases of $F$-mediated dark matter (DM) involve frozen-in keV-scale scalar DM, which may produce x-ray lines, and frozen-out TeV-scale fermionic DM. 
\end{abstract}

\maketitle

\tableofcontents

\section{Introduction}
\label{sec:intro}
Motivation for physics beyond the Standard Model (SM) typically falls into two categories. 
One is purely theoretical, often involving the unification of symmetries, such as supersymmetry, left-right symmetry and Grand Unified Theories. 
The other is experimentally driven, in which one tries to resolve problems or anomalies in the SM. 
This has led to, for instance, neutrino mass models, dark matter candidates, baryogenesis mechanisms and explanations of various flavour anomalies.

This dual approach to new physics has inspired a diverse range of models and phenomenological studies. 
It preferences depth rather than breadth of analysis: particularly compelling new physics models, such as the type-I seesaw mechanism, 
have been studied in extraordinary detail. 
Given the advanced state of the community and the recent lack of discoveries of new fundamental fields, it seems worthwhile to expand the search to include simple but generally less well-motivated possibilities. 
One must remember that the muon, for instance, was a great surprise at the time of its discovery, a seemingly unnecessary particle. 
Nobody knows how Nature will turn out. 
\\

There is only a small number of fermions which can have renormalisable interactions with SM fields. 
Most are already well-known, including the seesaw fermions and vector-like leptons and quarks. 
A weak doublet fermion with hypercharge $-3/2$, which we call $F$, has been almost entirely neglected in the literature. 
This field has previously been studied in only a handful of papers: aspects of collider phenomenology, electroweak (EW) precision data and lepton flavour physics were studied in Refs. \cite{delAguila:2008pw,deBlas:2013gla,Altmannshofer:2013zba,Ma:2014zda,Biggio:2014ela,Bizot:2015zaa}, while in Ref. \cite{Okada:2015hia,Kumar:2020web} the fermion $F$ was one of several new fields introduced to address problems in the SM. 
On one hand, this is understandable: it is not a dark matter candidate, nor can it explain neutrino masses (except as part of a three-loop model \cite{Okada:2015hia}), nor any of the various fine-tuning problems of the SM, not any of the recent flavour anomalies (except, again, as part of a larger model \cite{Kumar:2020web}). 
On the other hand, as we will show, adding this field to the SM via a $(\overline{F} \tilde{H} e_R + h.c.)$ term leads to rich phenomenology, including various lepton flavour violating processes and corrections to $Z$-boson decays. 
Moreover, it can be a mediator from the SM to simple but varied dark sectors. 
These novel DM candidates provides an additional motivation to study the $F$, and more generally to consider underappreciated extensions of the SM.

In Section \ref{sec:model}, we will introduce the model and perform a spurion analysis. 
Since the $F$ must be heavy, we will then derive the leading order WCs in the EFT description of the model. 
Having outlined the model, its phenomenology will be addressed in Section \ref{sec:pheno}, outlining the how the $F$ could have detectable effects in leptonic and electroweak scale observables. 
This is a more extensive analysis than the few previously performed in the literature. 
We turn to dark matter in Section \ref{sec:DM} and consider the most minimal scenarios which can provide a DM candidate via the `$F$-portal'.

\section{The Model and its effective description}
\label{sec:model}

\subsection{Model and spurion analysis}
We consider a single\footnote{One could imagine, in analogy with the SM, that there are several generations of $F$: this will be briefly discussed in Section \ref{ssec:severalgens}. } pair of chiral fermions added to the SM, $F_L$ and $F_R$, which are colour singlets and weak doublets with hypercharge $-3/2$. 
From now on, we write $F = F_L + F_R$. 
Since $F$ has both left- and right-handed components, the SM gauge symmetry remains anomaly-free. 
The Lagrangian is
\begin{equation}
    \mathcal{L} = \mathcal{L}_\text{SM} + \overline{F} (i \slashed{D} - M_F) F - (\overline{F} y_F \tilde{H} e_R + h.c.) \, ,
    \label{eq:Flagrangian}
\end{equation}
where $\tilde{H} = i \sigma_2 H^*$ and the covariant derivative is defined with a positive sign, i.e. $D_\mu F \equiv [\partial_\mu + (-3/2)ig_1 B_\mu + (1/2) g_2 \sigma^A W^A_\mu]F$. 
After electroweak (EW) symmetry breaking, the $F$ has singly- and doubly-charged components, $F = (F^- ~ F^{--})^T$.

The mass $M_F$ is taken to be real, which can be achieved by an appropriate rephasing of $F_L$ or $F_R$. 
Similarly, the $1 \times 3$ vector of Yukawa couplings, $y_F$, can be made real by rephasings of the $e_R$, $\mu_R$ and $\tau_R$ (the charged lepton Yukawas are kept real by rephasings of the lepton doublets). 
Consequently, only four real parameters are introduced in the model: $M_F$, $y_{Fe}$, $y_{F\mu}$ and $y_{F\tau}$, where $y_{F\alpha}$ denotes the coupling of the $F$ to the charged lepton flavour $\alpha$. 
In this sense, it is a very minimal\textemdash and hence very predictive\textemdash extension of the SM.

There is an enhanced flavour symmetry which results from introducing $F$. 
The SM itself contains an approximate $U(3)^5$ flavour symmetry, in which each of the five SM fermion species $\psi=l_L, e_R, q_L, u_R, d_R$ transforms as $\psi \to V_\psi \psi$ under the associated $U(3)_\psi$. 
It is useful to pretend that this symmetry is preserved by the Yukawa couplings if the Yukawas are interpreted as spurions which transform non-trivially under the symmetry. 
For instance, the charged lepton Yukawa interaction, $\mathcal{L} \supset -\overline{e_R} y_e H^\dagger l_L + h.c.$, is invariant given the transformation $y_e \to V_e y_e V_l^\dagger$. 
Including $F$, the SM flavour symmetry is extended by a $U(1)_F$, under which $F \to e^{i\alpha_F} F$. 
The new Yukawa interaction, $(\overline{F} y_F \tilde{H}e_R + h.c.)$, therefore respects the $U(3)^5 \times U(1)_F$ symmetry as long as $y_F$ has the transformation
\begin{equation}
    y_F \to e^{i\alpha_F} y_F V_e^\dagger \, .
    \label{eq:yF}
\end{equation}
It is convenient to define
\begin{equation}
    \epsilon_F \equiv \frac{v}{\sqrt{2} M_F} y_F \, .
    \label{epsilon}
\end{equation}
Clearly, $\epsilon_F$ transforms like $y_F$ under the flavour symmetry, however it also counts suppression by powers of $M_F$. 
This is useful when rewriting Eq. \eqref{eq:Flagrangian} in terms of the SMEFT. 
Using the properties of the flavour symmetry and counting powers of $M_F$ will allow us to deduce the form of the Wilson coefficients (WCs) which are generated by integrating out the heavy $F$.


\subsection{Effective description of the model}
It is already known that if $F$ exists, it must be heavier than the EW scale. 
If $M_F < m_Z/2$, the field would have been discovered in $Z$ decays at LEP \cite{ALEPH:2005ab}. 
This statement is independent of the size of $y_F$: the $Z \to F \bar{F}$ decay is a consequence of the EW charge of $F$. 
LHC searches have set a lower bound of a few hundred GeV \cite{Altmannshofer:2013zba,Ma:2014zda}. 
Moreover, as we will see in Section \ref{sec:pheno}, strong bounds can be obtained from other observables which impose that the three entries of $\epsilon_F$, defined in Eq. \eqref{epsilon}, must be $\ll 1$. 
It is therefore self-consistent to use an effective field theory (EFT) description of the model, and such an approach is often the most convenient. 
In particular, it makes power counting straightforward and it provides a common framework through which one can compare various different new physics models.

Taking the $F$ to be heavier than the EW scale, we can integrate it out at its mass scale and derive the corresponding SMEFT Lagrangian. 
Considering only operators up to dimension-6, its general form is
\begin{equation}
    \mathcal{L}_\text{SMEFT} = \mathcal{L}_\text{SM} + \mathcal{L}_\text{SMEFT} = \mathcal{L}_\text{SM} + \frac{\sqrt{2}}{v} c^W Q_W + \frac{2}{v^2} \sum \limits_i c^i \, Q_i \, .
\end{equation}
The single dim-5 operator, $Q_W$, is the Weinberg operator \cite{Weinberg:1979sa}, while at dim-6, we use the Warsaw basis of operators, $Q_i$, defined in Ref. \cite{Grzadkowski:2010es} after the earlier work of Ref. \cite{Buchmuller:1985jz}. 
The $c^i$ are the dimensionless WCs normalised by the EW scale.

The brief spurion analysis performed in the previous section aids an understanding of the general form of the WCs. 
The Lagrangian \eqref{eq:Flagrangian} is invariant under the $U(3)^5 \times U(1)_F$ flavour symmetry given the transformation \eqref{eq:yF}, therefore the EFT Lagrangian must similarly be invariant under the symmetry. 
In a (non-redundant) basis of operators, such as the Warsaw basis, each individual WC $c^i$ must transform in such a way that $c^i Q_i$ is invariant under the flavour symmetry. 
In particular, since every $Q_i$ is invariant under $U(1)_F$, because no SM field carries a $U(1)_F$ charge, each $c^i$ must also be invariant. 
All WCs must also have the correct power of $\epsilon_F$ in order to correspond to the dimension of the operator.

Let us start with the dim-5 Weinberg operator. 
This must have a single power of $\epsilon_F$. 
However, this is not $U(1)_F$-invariant, thus we can immediately see that non-zero $c^W$ cannot be generated.\footnote{This can equally be seen from the fact that the interactions of the $F$ do not violate lepton number, given an appropriate lepton number assignation of $F$.} 
At dim-6, we have two powers of $\epsilon_F$. 
From the transformation in Eq. \eqref{eq:yF}, we know only that each WC must have one power of $\epsilon_F$ or $\epsilon_F^T$ and one power of $\epsilon_F^*$ or $\epsilon_F^\dagger$ to be $U(1)_F$-invariant. 
The precise combination depends on the operator and on any SM couplings which are also present in the WC. 
Note that this discussion is independent of the number of families of $F$.

Integrating out the $F$ at tree-level up to dim-6, we find two operators with non-zero WCs,
\begin{equation}
    \mathcal{L}_\text{SMEFT}^\text{tree} = - \frac{2}{v^2} \frac{\epsilon_F^\dagger \epsilon_F}{2} Q_{He}  + \frac{2}{v^2} \left[ \frac{y_e^\dagger \epsilon_F^\dagger \epsilon_F}{2 } Q_{eH} + h.c. \right] \, ,
\end{equation}
i.e. the tree-level WCs of $Q_{He,\alpha \beta} = (\overline{e_{R\alpha}} \gamma_\mu e_{R\beta})(H^\dagger i \overleftrightarrow{D^\mu}H)$ and $Q_{eH,\alpha \beta} = (\overline{l_{L\alpha}} H e_{R\beta})(H^\dagger H)$ (see \cite{Grzadkowski:2010es} for the full list of dim-6 operators) are
\begin{align}
    &c^{He}_{\ab} = - \frac{1}{2} (\epsilon_F^\dagger \epsilon_F)_{\ab}&
    &c^{eH}_{\ab} = \frac{1}{2} (y_e^\dagger \epsilon_F^\dagger \epsilon_F)_{\ab}\, .&
    \label{treeWCs}
\end{align}
This is in agreement with \cite{delAguila:2008pw,deBlas:2017xtg}, and is consistent with the above discussion of flavour symmetries. 
Many additional SMEFT WCs are induced at one-loop leading log order via RGEs. 
These RGEs have been comprehensively computed in \cite{Jenkins:2013zja,Jenkins:2013wua,Alonso:2013hga}. 
Although they will prove to be unimportant for phenomenology, for completeness we list here the WCs generated at this order, evaluated at the EW scale, after running down from $M_F$: 
\begin{align}
	&c^H = -4\lambda \, \text{tr}[y_e y_e^\dagger \epsilon_F^\dagger \epsilon_F] \, L &
	&c^{H\square} = - \frac{1}{3} g_1^2 \, \text{tr}[\epsilon_F^\dagger \epsilon_F] \, L  & \notag \\
	&c^{HD} = - \frac{4}{3} g_1^2 \, \text{tr}[\epsilon_F^\dagger \epsilon_F] \, L &
	&c^{Hl(1)}_{\ab} = \frac{1}{3} g_1^2 \, \text{tr}[\epsilon_F^\dagger \epsilon_F] \delta_{\ab} \, L & \notag \\
	&c^{dH}_{ij} = 2\, \text{tr}[y_e y_e^\dagger \epsilon_F^\dagger \epsilon_F ] (y_d^\dagger)_{ij} \, L &
	&c^{uH}_{ij} = 2\, \text{tr}[y_e y_e^\dagger \epsilon_F^\dagger \epsilon_F] (y_u^\dagger)_{ij} \, L & \notag \\
	&c^{Hq(1)}_{ij} = -\frac{1}{9} g_1^2 \, \text{tr}[\epsilon_F^\dagger \epsilon_F] \delta_{ij} \, L &
	&c^{Hu}_{ij} = - \frac{4}{9} g_1^2 \, \text{tr}[\epsilon_F^\dagger \epsilon_F] \delta_{ij} \, L & \notag \\
	&c^{Hd}_{ij} = \frac{2}{9} g_1^2 \, \text{tr}[\epsilon_F^\dagger \epsilon_F] \delta_{ij} \, L &
	&c^{ee}_{\alpha \beta \gamma \delta} = - \frac{1}{6} g_1^2 \left[ \delta_{\alpha \beta} ( \epsilon_F^\dagger \epsilon_F)_{\gamma \delta} + ( \epsilon_F^\dagger \epsilon_F)_{\alpha \beta} \delta_{\gamma \delta} \right] \, L & \notag \\
	&c^{ed}_{\ab ij} = - \frac{1}{9} g_1^2 ( \epsilon_F^\dagger \epsilon_F)_{\alpha \beta} \delta_{ij} \, L &
	&c^{eu}_{\ab ij} = (\epsilon_F^\dagger \epsilon_F)_{\alpha \beta} \left[ \frac{2}{9} g_1^2 \delta - y_u y_u^\dagger \right]_{ij} \, L  & \notag \\
	&c^{le}_{\ab \cd} = - \frac{1}{6} g_1^2 \delta_{\alpha \beta} (\epsilon_F^\dagger \epsilon_F)_{\gamma \delta} \, L &
	&c^{qe}_{ij\ab} = \frac{1}{2} ( \epsilon_F^\dagger \epsilon_F)_{\ab} \left[ \frac{1}{9} g_1^2 \delta + y_u^\dagger y_u \right]_{ij}  \, L \, ,&
\end{align}
where $L \equiv \ln (M_F/v)/(16\pi^2)$. 
We have systematically neglected terms suppressed by powers of the charged lepton and down quark Yukawas, $y_e$ and $y_d$, compared to terms proportional to powers of the EW gauge couplings $g_1$ and $g_2$, since $g_{1,2}^2 \gg y_b^2$.

The EW dipole WCs, $c^{eB}$ and $c^{eW}$, are relevant for phenomenology despite being induced neither at tree-level nor at one-loop leading-log level. 
They determine the rate of radiative charged lepton decays and of corrections to charged lepton electric and magnetic dipole moments. 
Several steps are required to obtain the correct dipole WCs in EFT, see e.g. \cite{Aebischer:2021uvt,Coy:2021hyr}. First, one-loop matching at the scale $M_F$ onto the EW dipole operators gives
\begin{align}
    &c^{eB}_{\alpha \beta} = \frac{g_1}{48\pi^2} (y_e^\dagger \epsilon_F^\dagger \epsilon_F)_{\alpha \beta} &
    &c^{eW}_{\alpha \beta} = \frac{g_2}{128\pi^2} (y_e^\dagger \epsilon_F^\dagger \epsilon_F)_{\alpha \beta} \, , &
\end{align}
where we cross-checked the necessary loop integrals with Package-X \cite{Patel:2015tea}. 
Neglecting the running of these operators, which is a two-loop effect, they match onto the electromagnetic dipole operator of the low-energy EFT, giving
\begin{equation}
    c^{e\gamma,\text{UV}}_{\alpha \beta} = c_w c^{eB} - s_w c^{eW} = \frac{5e}{384\pi^2} (y_\alpha^\dagger \epsilon_F^\dagger \epsilon_F)_{\alpha \beta}  \, .
\end{equation}
Secondly, one-loop matching of $Q^{He}$ onto the EM dipole operator at the EW scale gives a contribution of the same order \cite{Dekens:2019ept}, 
\begin{equation}
    c^{e\gamma,\rm EW}_{\alpha \beta} = \frac{e s_w^2}{12\pi^2} (y_e^\dagger c^{He})_{\alpha \beta} = - \frac{e s_w^2}{24\pi^2} (y_\alpha^\dagger \epsilon_F^\dagger \epsilon_F)_{\alpha \beta} \, ,
\end{equation}
with $s_w$ the sine of the weak-mixing angle. 
Finally, at the charged lepton mass scale, the contribution to dipole observables from loops involving four-lepton operators (which $Q_{He}$ matches onto at the EW scale) corresponds to \cite{Aebischer:2021uvt}
\begin{equation}
    c^{e\gamma,\text{eff}}_{\alpha \beta} = - \frac{e (1 - 2s_w^2)}{32\pi^2} ( y_\alpha^\dagger \epsilon_F^\dagger \epsilon_F)_{\alpha \beta}  \, .
\end{equation}
Summing these pieces, the total observable EM dipole WC is
\begin{align}
    c^{e\gamma,\text{obs}}_{\alpha \beta} =  c^{e\gamma,\text{UV}}_{\alpha \beta} + c^{e\gamma,\rm EW}_{\alpha \beta} + c^{e\gamma,\text{eff}}_{\alpha \beta} = \frac{(-7+8s_w^2)e}{384\pi^2} (y_\alpha^\dagger \epsilon_F^\dagger \epsilon_F)_{\alpha \beta} \, .
    \label{eq:cegfull}
\end{align}
This disagrees with the only previous calculation in the literature \cite{Biggio:2014ela}, although the UV parts agree.

Having derived the tree-level, one-loop leading-log and dipole WCs of the SM$+F$ model, we can now turn to its phenomenology.

\section{Phenomenological analysis}
\label{sec:pheno}
Many phenomenological studies of the SMEFT have previously been performed, see e.g. \cite{Crivellin:2013hpa,Falkowski:2014tna,Berthier:2015oma,Feruglio:2015gka,Falkowski:2015krw,Falkowski:2017pss,Frigerio:2018uwx,Coy:2021hyr}. 
Here in particular we use the bounds compiled in table 7 of \cite{Coy:2021hyr}. 
The great deal of work already done to bound WCs of the SMEFT, and the easy and general applicability of these results, indeed provides additional motivation for studying new physics models within the framework of EFT.

Since all WCs depend on $(\epsilon_F^\dagger \epsilon_F)$, all bounds will be on this combination of parameters. 
For only one family of $F$, $(\epsilon_F^\dagger \epsilon_F)_{ab} = y_{Fa} y_{Fb} v^2/(2M_F^2)$. 
However, we keep in mind the possibility that there may be several generations of $F$ and therefore express the bounds in matrix form.

\subsection{EW scale observables}
The most constraining EW scale observables on new physics in the lepton sector typically come from corrections to EW input parameters and from modifications to the decays of EW-scale fields. 

\subsubsection*{EW input parameters} 
The $Z$-boson mass, $m_Z$, the Fermi constant, $G_F$, and the electromagnetic fine-structure constant, $\alpha$, are the precisely-measured inputs used to make SM predictions for a range of other observables. 
None of these are modified by the two dim-6 operators generated at tree-level, $Q_{He}$ and $Q_{eH}$. 
Although an additional muon decay channel is induced at tree-level, since $Q_{\substack{He\\ e\mu}}$ leads to $\mu_R \to e_R \bar{\nu}_{L\alpha} \nu_{L\alpha}$, this does not interfere with the SM decay, $\mu_L \to e_L \bar{\nu}_{Le} \nu_{L\mu}$, and thus the correction to $G_F$ is $\mathcal{O}(\epsilon_F^4)$.\footnote{One also needs to compute the EFT up to dim-8 to correctly find the shift in muon decay, since the interference of a dim-8 WC with the SM is also $\mathcal{O}(\epsilon_F^4)$. } 
The changes to the input parameters are therefore subdominant compared to other observables, many of which arise at tree-level and $\mathcal{O}(\epsilon_F^2)$, which will shortly be discussed. 
We note also that for $|\epsilon_F| \lesssim 1$ the model induces a negligible shift in $m_W$ (the largest effect is either $\mathcal{O}(\epsilon_F^4)$ or $\mathcal{O}(\epsilon_F^2 L)$) and cannot explain the recent measurement reported by CDF \cite{CDF:2022hxs}.

\subsubsection*{Higgs boson decays}
We now turn to the decays of the Higgs and $Z$ bosons. 
While flavour-conserving Higgs decays to leptons are not yet precisely measured \cite{ParticleDataGroup:2020ssz}, better constraints come from flavour-violating decays. 
The width of these decays is
\begin{equation}
    \Gamma(h \to \ell_\alpha \ell_\beta) \equiv \Gamma(h \to \ell_\alpha^+ \ell_\beta^-) + \Gamma(h \to \ell_\alpha^- \ell_\beta^+) \simeq \frac{m_h (m_\alpha^2 + m_\beta^2)}{8\pi v^2} | (\epsilon_F^\dagger \epsilon_F)_{\alpha \beta}|^2  \, ,
\end{equation}
using $|(\epsilon_F^\dagger \epsilon_F)_{\alpha \beta}| = |(\epsilon_F^\dagger \epsilon_F)_{\beta \alpha}|$. 
The current (future) bounds \cite{ATLAS:2019old,CMS:2021rsq} (\cite{Banerjee:2016foh}) give the limits
\begin{align}
    &|(\epsilon_F^\dagger \epsilon_F)_{\mu e}| \lesssim 0.53 (0.26) \, , &
    &|(\epsilon_F^\dagger \epsilon_F)_{\tau e}| \lesssim 0.19 (0.071) \, , &
    &|(\epsilon_F^\dagger \epsilon_F)_{\tau \mu}| \lesssim 0.16 (0.071) \, . &
\end{align}

\subsubsection*{$Z$-boson decays}
Stronger bounds come from $Z$-boson decays. 
The experimental limits 
are better; moreover $c^{eH}$, which generates Higgs decays, is suppressed by the small charged lepton Yukawa couplings, while $c^{He}$, which generates $Z$ decays, does not have this suppression, cf. Eq. \eqref{treeWCs}. 
Flavour-violating decays occur with a rate
\begin{equation}
     \Gamma(Z \to \ell_\alpha \ell_\beta) \equiv \Gamma(Z \to \ell_\alpha^+ \ell_\beta^-) + \Gamma(Z \to \ell_\alpha^- \ell_\beta^+) \simeq \frac{m_Z^3}{12\pi v^2} |(\epsilon_F^\dagger \epsilon_F)_{\alpha \beta}|^2 \, .
\end{equation}
The best current limits come from the LHC \cite{ATLAS:2021bdj,ATLAS:2022uhq} , 
which gives
\begin{align}
    &|(\epsilon_F^\dagger \epsilon_F)_{e\mu}| \lesssim 1.4 \times 10^{-3} \, , &
    &|(\epsilon_F^\dagger \epsilon_F)_{e\tau}| \lesssim 6.2 \times 10^{-3} \, , &
    &|(\epsilon_F^\dagger \epsilon_F)_{\mu\tau}| \lesssim 7.0 \times 10^{-3} \, . & 
\end{align}
Unlike for the Higgs decays, flavour-conserving $Z$-boson decays also provide a stringent constraint. 
This is due to the extremely precise measurement of the various $Z$ decay channels at LEP \cite{ALEPH:2005ab,ParticleDataGroup:2020ssz}. 
The bounds obtained in \cite{Coy:2021hyr} specified to this model lead to
\begin{align}
    &|(\epsilon_F^\dagger \epsilon_F)_{ee}| \lesssim 1.1 \times 10^{-3}  \, , &
    &|(\epsilon_F^\dagger \epsilon_F)_{\mu \mu}| \lesssim 2.3 \times 10^{-3} \, ,&
    &|(\epsilon_F^\dagger \epsilon_F)_{\tau \tau}| \lesssim 4.3 \times 10^{-3} \, . & \label{Zeebound}
\end{align}
Similarly, the bound from $Z$-boson decays into neutrinos, parameterised by the effective number of neutrinos, $N_\nu$, gives
\begin{equation}
    \text{tr}[\epsilon_F^\dagger \epsilon_F] \lesssim 0.010 \, .
\end{equation}
This bound is however weaker than the sum of the constraints on individual flavours from $Z \to \ell_\alpha^+ \ell_\alpha^-$, listed in Eq. \eqref{Zeebound}.

\subsubsection*{Weak-mixing angle}
Finally, one can obtain a competitive bound from the measurement of $s_w^2$. 
As explained in \cite{Coy:2021hyr}, the WC $c^{He}$ is bounded by this observable since it affects various asymmetries of $Z$-boson decays which LEP uses to extract the value of $s_w^2$ \cite{ALEPH:2005ab}. 
The estimated bound corresponds to \cite{Coy:2021hyr}
\begin{equation}
    -1.7 \times 10^{-3} \lesssim \text{tr}[c^{He}] \lesssim 1.1 \times 10^{-3} \, ,
\end{equation}
from which we obtain
\begin{equation}
    \text{tr}[\epsilon_F^\dagger \epsilon_F] \lesssim 3.4 \times 10^{-3} \, .
\end{equation}
This is about a factor of 3 stronger than the bound from $N_\nu$, and improves upon the bound on $|(\epsilon_F^\dagger \epsilon_F)_{\tau \tau}|$ given in Eq. \eqref{Zeebound}.

\subsection{Lepton mass scale observables}
Next we consider observables at the charged lepton mass scale. 
The most relevant of these fall into two categories: i) flavour-violating charged lepton transitions, and ii) dipole moments. 

\subsubsection*{Lepton decays to three charged leptons}
The rates for lepton to three lepton decays which violate flavour by one unit are
\begin{align}
    \Gamma(\ell_\alpha^- \to \ell_\beta^- \ell_\beta^+ \ell_\beta^-) &\simeq \frac{m_\alpha^5}{1536\pi^3 v^4} ( 1 - 4 s_w^2 + 12 s_w^4 ) |(\epsilon_F^\dagger \epsilon_F)_{\alpha \beta}|^2 \\
    \Gamma(\tau^- \to \ell_\beta^- \ell_\beta^+ \ell_\gamma^-) &\simeq \frac{m_\tau^5}{1536\pi^3 v^4} ( 1 - 4 s_w^2 + 8 s_w^4 ) |(\epsilon_F^\dagger \epsilon_F)_{\tau \gamma}|^2 \, ,
\end{align}
including the contribution from $c^{He}$ but neglecting the Yukawa-suppressed contribution from $c^{eH}$. 
The rate for $\ell_\alpha^- \to \ell_\beta^- \ell_\beta^+ \ell_\beta^-$ is also computed in \cite{Altmannshofer:2013zba}, and we find agreement. 
From the current (future) bounds on $\mu \to 3e$ \cite{Bellgardt:1987du} (\cite{Blondel:2013ia}), $\tau \to 3e$ \cite{Hayasaka:2010np} (\cite{Kou:2018nap}) and $\tau \to 3\mu$ \cite{Hayasaka:2010np} (\cite{Kou:2018nap}) branching ratios, we obtain the bounds
\begin{align}
    &|(\epsilon_F^\dagger \epsilon_F)_{\mu e}| \lesssim 24 (0.24) \times 10^{-7} \, , &
    &|(\epsilon_F^\dagger \epsilon_F)_{\tau e}| \lesssim 9.2 (1.2) \times 10^{-4} \, ,&
    &|(\epsilon_F^\dagger \epsilon_F)_{\tau \mu}| \lesssim 8.1 (1.0) \times 10^{-4} \, .&
\end{align}
From the current (future) bounds on $\tau^- \to e^- e^+ \mu^-$ \cite{Hayasaka:2010np} (\cite{Kou:2018nap}) and $\tau^- \to \mu^- \mu^+ e^-$ \cite{Hayasaka:2010np} (\cite{Kou:2018nap}), we find
\begin{align}
    &|(\epsilon_F^\dagger \epsilon_F)_{\tau \mu}| \lesssim 9.0 (1.2) \times 10^{-4} \, , &
   &|(\epsilon_F^\dagger \epsilon_F)_{\tau e}| \lesssim 11 (1.4) \times 10^{-4} \, &
\end{align}
These are stronger than the bounds from CLFV $Z$-boson decays, most notably in the $\mu \to e$ sector where the improvement is three orders of magnitude. 
This is due to the impressive experimental limits on muon and tau decays: the branching ratio of LFV $Z$-boson decays are bounded at the $\mathcal{O}(10^{-6})$ level, tau branching ratios are probed at $\mathcal{O}(10^{-8})$ while for the muon the precision is $\mathcal{O}(10^{-13})$.

At tree-level there are no charged leptons decays which violate flavour by two units, e.g. $\tau^- \to e^- e^- \mu^+$. 
Since the branching ratios are constrained to a similar level as those of decays which violate flavour by a single unit, the bounds from these processes are significantly weaker.

\subsubsection*{Radiative charged lepton decays}
The rate of radiative charged lepton decays is
\begin{equation}
    \Gamma( \ell_\alpha \to \ell_\beta \gamma) \simeq \frac{m_\alpha^3}{2\pi v^2} \left( |c^{e\gamma,\rm obs}_{\alpha \beta}|^2 + |c^{e\gamma,\rm obs}_{\beta \alpha}|^2 \right)
\end{equation}
From the current (future) limits on $\mu \to e \gamma$ \cite{TheMEG:2016wtm} (\cite{MEGII:2018kmf}), $\tau \to e \gamma$ \cite{Aubert:2009ag} (\cite{Kou:2018nap}) and $\tau \to \mu \gamma$ \cite{Aubert:2009ag} (\cite{Kou:2018nap}), this gives 
\begin{align}
    &|(\epsilon_F^\dagger \epsilon_F)_{\mu e}| \lesssim 2.6 (1.0) \times 10^{-5}  \, ,&
    &|(\epsilon_F^\dagger \epsilon_F)_{\tau e}| \lesssim 0.017 (0.005) \, ,&
    &|(\epsilon_F^\dagger \epsilon_F)_{\tau \mu}| \lesssim 0.019 (0.003) \, ,
\end{align}
using Eq. \eqref{eq:cegfull}. 
It is unsurprising that these bounds are weaker than those from $\ell \to 3\ell$ decays, since the experimental limits on such processes are comparable, but $\ell \to 3 \ell$ is induced at tree-level in this model while radiative decays only at one-loop.

\subsubsection*{$\mu \to e$ conversion in nuclei}
The rate of $\mu \to e$ conversion in nuclei due to the $F$ is \cite{Cirigliano:2009bz,Crivellin:2017rmk}
\begin{align}
    \Gamma_N &= \frac{m_\mu^5}{v^4} \left[ (1 - 4 s_w^2 ) V^p_N - V^n_N \right]^2 \left|(\epsilon_F^\dagger \epsilon_F)_{e\mu} \right|^2 \, ,
\end{align}
keeping only the contribution from $c^{He}$, since the contribution from $c^{eH}$ is relatively suppressed by a factor $m_{p,n}/v \simeq 0.004$ in the amplitude. 
This result is in agreement with \cite{Altmannshofer:2013zba}. 
The nucleus-dependent form factors $V^{p,n}_N$ are given in Table 1 of \cite{Kitano:2002mt}. 
The current bound from conversion in gold \cite{Bertl:2006up} gives
\begin{equation}
    \left|(\epsilon_F^\dagger \epsilon_F)_{e\mu} \right| \lesssim 3.0 \times 10^{-7} \, .
\end{equation}
This is the single strongest current bound on the model. 
For $y_{Fe} = y_{F\mu} = 1$, it corresponds to $M_F > 320$ TeV. 
The expected future bounds from conversions in aluminium \cite{Kuno:2013mha} and titanium \cite{Barlow:2011zza,Knoepfel:2013ouy}
are
\begin{equation}
    \left|(\epsilon_F^\dagger \epsilon_F)_{e\mu} \right| \lesssim ( 70,\, 5.0)  \times 10^{-10}  \, , ~ (\text{Al, Ti}) \, .
\end{equation}
The latter is the best expected future limit and corresponds to $M > 7.8$ PeV for $y_{Fe} = y_{F\mu} = 1$.

\subsubsection*{Dipole moments}
Finally we turn to dipole moments. 
This scenario generates a small negative shift in magnetic dipole moments of charged leptons at one loop, 
\begin{equation}
    \Delta a_\alpha = - \frac{(7-8s_w^2)}{48\pi^2} \frac{ m_\alpha^2}{v^2} ( \epsilon_F^\dagger \epsilon_F)_{\alpha \alpha} \, .
\end{equation}
This is the opposite direction to the putative anomaly in the magnetic moment of the muon \cite{Aoyama:2020ynm,Abi:2021gix} (notwithstanding the lattice results which put into question the existence of the anomaly \cite{Borsanyi:2020mff}). 
From this, we find the bound
\begin{equation}
    (\epsilon_F^\dagger \epsilon_F)_{\mu \mu} \lesssim 0.2 \, .
\end{equation}
Meanwhile, we consider a conservative bound on the magnetic moment of the electron, considering as a systematic uncertainty the discrepancy in measured values of the fine-structure constant in different atoms \cite{Parker:2018vye,Morel:2020dww}, $|\text{Re} c^{e\gamma,\text{obs}}_{ee} | \lesssim 3 \times 10^{-8}$ \cite{Coy:2021hyr}. 
This is compatible with $\epsilon_{Fe} \sim \mathcal{O}(1)$.

The electric dipole moment, by contrast, does not appear until the two-loop level, and only if there are at least two families of $F$. 
When there is a single fermion $F$, the entries of the $1 \times 3$ vector $y_{F}$ can be made real, as argued in Section \ref{sec:model}, and hence there are no new CPV interactions. 
Assuming that $y_F$ is a $n_F \times 3$ matrix with some complex entries, where $n_F$ is the number of generations of $F$, the contribution to the EDM of the charged leptons can be estimated using a spurion analysis similar to the one performed for the type-I seesaw mechanism in \cite{Smith:2017dtz,Coy:2018bxr}. 
The estimate is
\begin{align}
    |d_\alpha| &\sim \frac{4e m_\alpha}{(16\pi^2)^2 v^2} \text{Im}\left[[\epsilon_F^\dagger \epsilon_F, \epsilon_F^\dagger \epsilon_F y_e y_e^\dagger \epsilon_F^\dagger \epsilon_F]_{\alpha \alpha} \right] \notag \\ \Rightarrow |d_e| &\sim 5.6 \times 10^{-30} \, \text{Im}\left[(\epsilon_F^\dagger \epsilon_F)_{e\tau} (\epsilon_F^\dagger \epsilon_F)_{\tau \mu} (\epsilon_F^\dagger \epsilon_F)_{\mu e} \right] e \text{ cm} \, .
\end{align}
Given the strong bounds from flavour-violation outlined above, the EDM is clearly below not only the experimental upper limit of $1.1 \times 10^{-29} e$ cm \cite{ACME:2018yjb} but also the estimated SM prediction of $\mathcal{O}(10^{-38})e$ cm \cite{Pospelov:2013sca,Ghosh:2017uqq}.

\subsection{Comparison of rates}
As stated previously, the SM$+F$ model is highly predictive as there are few new parameters. 
This allows us to directly correlate the rates of different processes, with
\begin{align}
    \text{BR}(\mu \to e\gamma) &= 2.9\, \text{BR}(h \to e \mu) = 4.8 \times 10^{-3} \text{BR}(Z \to e \mu) \\
    &= 3.6 \times 10^{-3} \text{BR}(\mu^- \to e^- e^+ e^-) = 2.8 \times 10^{-5} \text{BR}(\mu \text{Ti} \to e \text{Ti}) \notag \\
    \text{BR}(\tau \to \ell_\alpha \gamma) &= 1.9 \times 10^{-3} \text{BR}(h \to \ell_\alpha \tau) = 8.6 \times 10^{-4} \text{BR}(Z \to \ell_\alpha \tau) \\
    &= 3.6 \times 10^{-3} \text{BR}(\tau^- \to \ell_\alpha^- \ell_\alpha^+ \ell_\alpha^-) = 5.1 \times 10^{-3} \text{BR}(\tau^- \to \ell_\alpha^- \ell_\beta^+ \ell_\beta^-) \notag 
\end{align}
for $\alpha = e,\mu$ and $\beta \neq \alpha, \tau$. 
These relations hold even for several generations of $F$. 
If in the future one of these CLFV processes were observed, we would therefore have strict predictions for the expected rates of others which violate flavour in the same way. 
This would allow us to determine whether or not the $F$ was responsible for the new physics. 
Since all the observables discussed depend only on combinations of $\epsilon_F$, measurements would allow us to determine the values of the entries of $\epsilon_F$ but not the mass of the $F$.


\subsection{Summary and plots}
\begin{figure}[t!]
    \centering
    \includegraphics[width=0.75\textwidth]{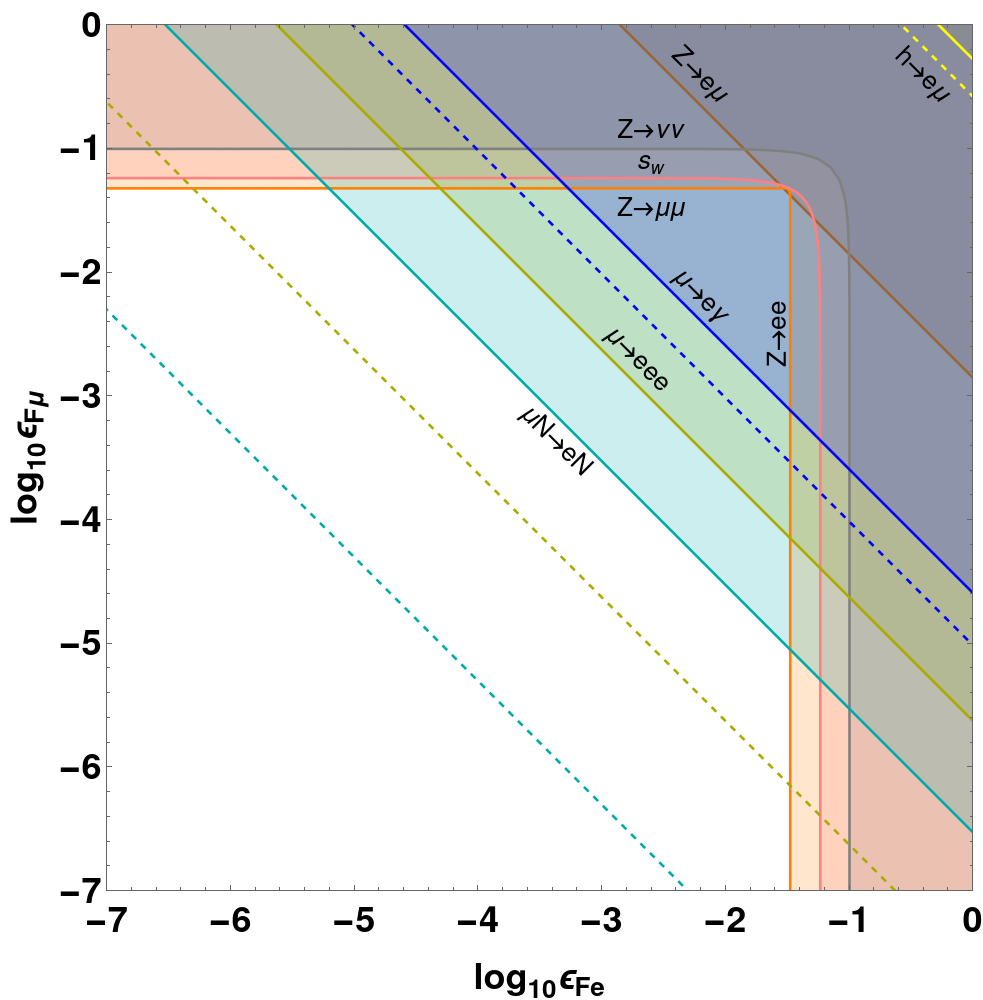}
    \caption{Bounds on the model in the $e-\mu$ sector, setting $\epsilon_{F\tau} = 0$. 
    Flavour-violating constraints come from $\mu \to e$ conversion (teal), $\mu \to 3e$ (mustard), $\mu \to e \gamma$ (blue), $Z \to e \mu$ (brown) and $h \to e \mu$ (yellow). Flavour-conserving constraints come from $Z \to \ell_\alpha^+ \ell_\alpha^-$ (orange), $s_w$ (pink), and $Z \to \nu \nu$ (grey). Solid lines correspond to existing bounds, dashed ones to expected future bounds. }
    \label{fig:emu}
\end{figure}
\begin{figure}[t!]
    \centering
    \includegraphics[width=0.75\textwidth]{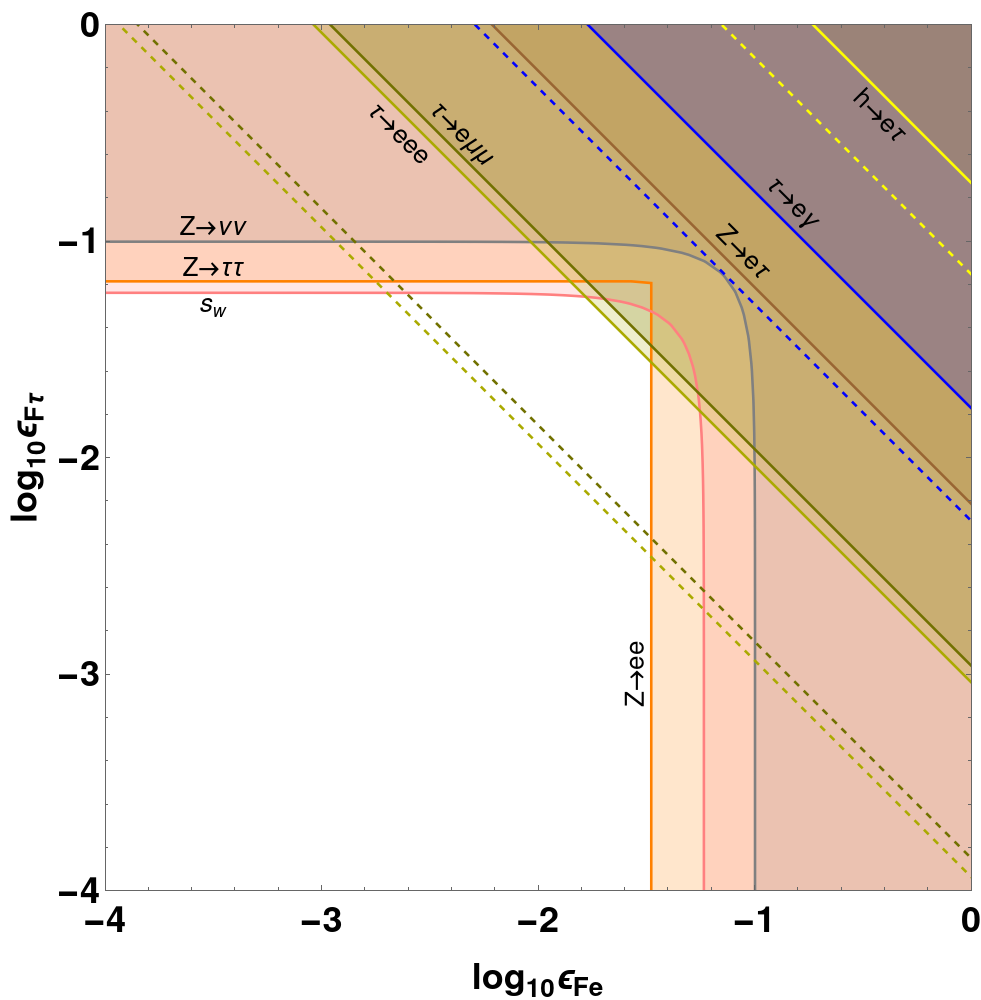}
    \caption{Bounds on the model in the $e-\tau$ sector, setting $\epsilon_{F\mu} = 0$. 
    The colour scheme is as in Fig. \ref{fig:emu}, up to $\mu \leftrightarrow \tau$, additionally the constraints from $\tau^- \to e^- \mu^+ \mu^-$ are in dark mustard. }
    \label{fig:etau}
\end{figure}
\begin{figure}[t!]
    \centering
    \includegraphics[width=0.75\textwidth]{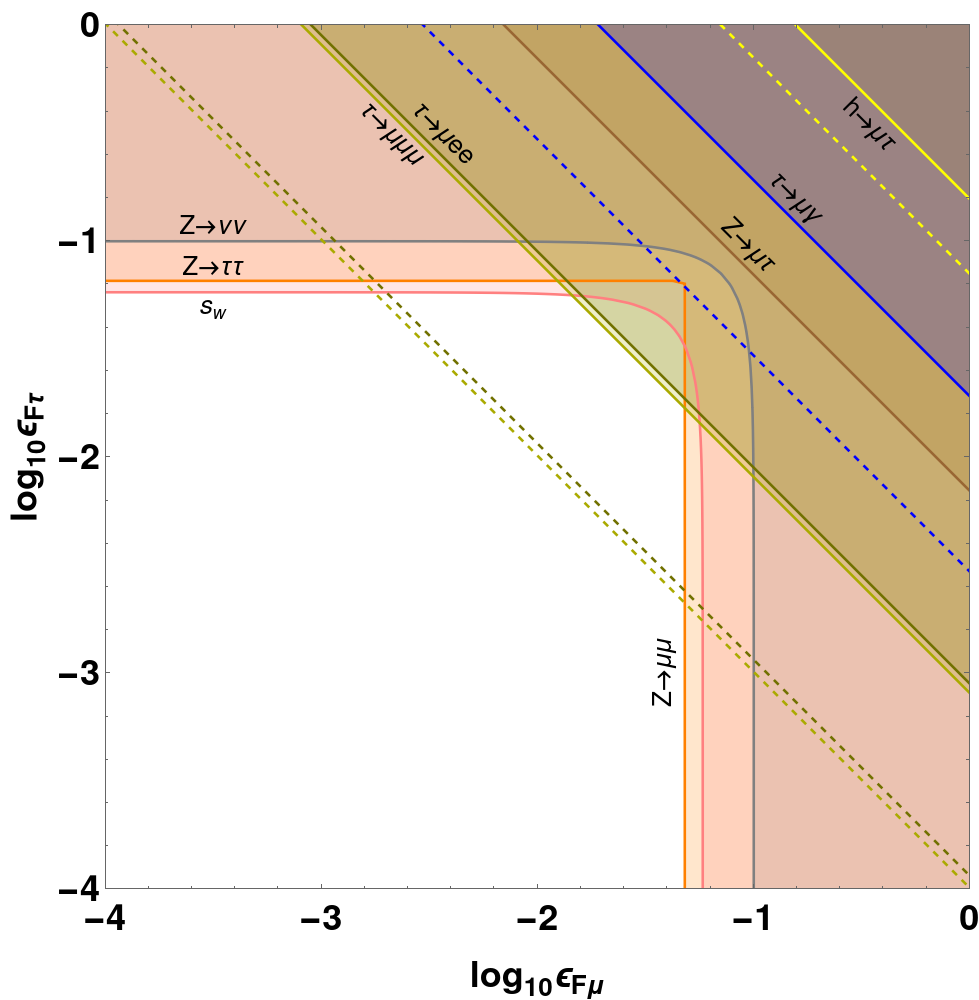}
    \caption{Bounds on the model in the $\mu-\tau$ sector, setting $\epsilon_{Fe} = 0$. 
    The colour scheme is as in Fig. \ref{fig:etau}, up to $e \leftrightarrow \mu$.}
    \label{fig:mutau}
\end{figure}

The results are summarised in Figs. \ref{fig:emu}-\ref{fig:mutau}. 
In each case, we consider the possibility that the $F$ is coupled to two flavours of leptons but not to the third, i.e. two entries of $\epsilon_F$ are non-zero while the third is set to zero. 
Then the small number of parameters in the model allows us to directly compare flavour-conserving and flavour-violating observables. 
We benefit from the analysis of \cite{Coy:2021hyr}, which can be applied to any model of new physics in the lepton sector, including this $F$ model. 
Therefore, the plots presented here can be directly compared with those in \cite{Coy:2021hyr}, assuming equal values of the $\epsilon$ for each different model.

Fig. \ref{fig:emu} shows the $e-\mu$ sector. 
It is clear that when $\epsilon_{Fe} \sim \epsilon_{F\mu}$, the flavour-violating observables are much more constraining than flavour-conserving ones. 
The strongest current and expected future limits both come from $\mu \to e$ conversion, given by the solid and dashed teal lines, respectively. 
When one of $\epsilon_{Fe}$ or $\epsilon_{F\mu}$ becomes very small, the flavour-conserving bounds take over since all $\mu \to e$ processes have rates proportional to $\epsilon_{Fe}^2 \epsilon_{F\mu}^2$. 
In particular, flavour-conserving $Z$-boson decays bound $\epsilon_{Fe,\mu} \lesssim 10^{-1.5}$, which corresponds to $M_F > 5$ TeV for $y_{Fe,\mu} = 1$.

In Figs. \ref{fig:etau} and \ref{fig:mutau}, we see that flavour-violating and flavour-conserving bounds are presently of similar strength: both bound $\epsilon_{F\alpha} \lesssim 10^{-1.5}$. 
However, the prospects for improved measurements of flavour-violating $\tau$ decays to three charged leptons at Belle II \cite{Kou:2018nap} means that the limits are expected to improve by nearly an order of magnitude. 
Given this, improvements in the measurement of flavour-conserving $Z$-boson decays or of the weak-mixing angle would provide a useful complementary test of the model.

One also observes that flavour-violating Higgs and $Z$-boson decays prove to be unimportant for phenomenology, as do radiative charged lepton decays. 
The relatively mild constraints from Higgs and $Z$ decays is due to the poorer experimental bounds on their decays compared to the extremely precise measurements of charged lepton decays. 
Meanwhile, the radiative decays give weaker limits compared to $\ell \to 3 \ell$ decays since the latter proceeds at tree-level while the former is loop-suppressed.

Two of the most notable aspects of the $F$ phenomenology are i) that the bounds from $\mu \to 3e$ significantly exceed those from $\mu \to e \gamma$, and ii) the lack of a competitive bound from $m_W$. 
Consider the first of these points. 
The current experimental limits on the two decays are rather similar, $\text{BR}(\mu \to 3e) < 10^{-12}$ and $\text{BR}(\mu \to e \gamma) < 0.42 \times 10^{-12}$, thus the simplest expectation is that they would lead to similar bounds on the parameter space. 
In this model, however, $\mu \to 3e$ proceeds at tree-level with no chirality flip, while $\mu \to e \gamma$ is a one-loop process and has a small charged lepton Yukawa suppression, so the former decay leads to a much stronger constraint. The story is in fact the same for the type-III seesaw (see e.g. \cite{Coy:2021hyr}). 
In many other models, however, $\mu \to 3e$ occurs at loop-level and/or $\mu \to e \gamma$ avoids its usual Yukawa suppression (which it might if, for instance, there are new scalars which have $\mathcal{O}(1)$ couplings to charged leptons), thus the bounds from the two observables are more similar. 
This comparison is particularly relevant given the anticipated four orders of magnitude improvement to the experimental sensitivity to $\mu \to 3e$ at the Mu3e experiment \cite{Blondel:2013ia}.

The second point, the fact that the model modifies various EW-scale observables but negligibly impacts $m_W$,\footnote{This is perhaps a nice contrast to the plethora of recent models which sought to explain the CDF anomaly \cite{CDF:2022hxs}.} 
arises from the fact that the new state couples to the lepton singlet. 
When new physics couples to the lepton doublets, they typically interfere with SM muon decay at tree-level, as explained at the beginning of this section. 
This modifies not only $G_F$, but also indirectly affects $m_W$, $s_w$ and $Z \to \ell_\alpha^+ \ell_\alpha^-$, see e.g. \cite{Coy:2021hyr} and the four models studied therein. 
Conversely, the SM$+F$ model changes $Z$-boson couplings, and hence both the measurement of $s_w$ and $Z \to \ell_\alpha^+ \ell_\alpha^-$, but negligibly impacts muon decay and therefore $m_W$. 
Future deviations in the former observables but not the latter would therefore provide evidence for the presence of the $F$ while at the same time excluding a large number of recently studied models.

\subsection{Several generations of \texorpdfstring{$F$}{F}}
\label{ssec:severalgens}
Going beyond the simplest case, one could imagine that there may be more than one copy of $F$. 
In that case, the flavour-conserving and flavour-violating observables are in general not so tightly correlated.

The matrix $(\epsilon_F^\dagger \epsilon_F)$ is positive semi-definite, thus we have the Cauchy-Schwarz inequality, $|(\epsilon_F^\dagger \epsilon_F)_{\alpha \beta}| \leq \sqrt{(\epsilon_F^\dagger \epsilon_F)_{\alpha \alpha}(\epsilon_F^\dagger \epsilon_F)_{\beta \beta}}$. 
The case of a single $F$ therefore maximises flavour-violation as this inequality being saturated. 
When there are two generations of $F$, it is possible for two off-diagonals to be zero while all diagonal entries are non-zero. 
When there are three generations of $F$, it is possible to modify all flavour-conserving observables while forbidding flavour violation as each copy of $F$ can couple to a different family of SM leptons. In this case, all off-diagonals are zero while the diagonals of $(\epsilon_F^\dagger \epsilon_F)$ are non-zero. 

\section{\texorpdfstring{$F$}{F}-portal dark matter}
\label{sec:DM}
In the second part of this paper, we investigate the role that $F$ can play in communicating with the dark sector. 
Clearly, neither of the two components of $F$ can be dark matter: both are electrically charged and decay quickly due to the $\overline{F} \tilde{H} e_R$ interaction.
Here we search for minimal extensions of the SM+$F$ model which produce a viable DM candidate. 
First, we will classify the possible neutral particles which can be added and find that a SM singlet scalar could be a viable DM candidate, though the model requires several very small parameters. 
Then we will provide an example of a DM candidate in a two field extension where all new couplings may be $\mathcal{O}(1)$. 
Both cases intersect in different ways with the bounds obtained in Section \ref{sec:pheno}.


\subsection{Single field extensions}
Consider the possibility of adding a single additional field which has renormalisable interactions with the $F$ and gives a DM candidate. 
There are three candidate fields which are charged under the SM and one which is neutral. 

We begin with the fields charged under the SM gauge symmetry: i) a fermion triplet, $\Psi \sim \mathbf{3}_{-1}$, with $\overline{F}\sigma^A \tilde{H} \Psi^A$ and $\overline{l_L} \sigma^A H \Psi^A$ interactions, ii) a scalar doublet, $\Phi \sim \mathbf{2}_{1/2}$, with a $\overline{F} \tilde{\Phi} e_R$ interaction, and a scalar triplet, $\Delta \sim \mathbf{3}_{-1}$, with a $\overline{F} \sigma^A l_L \Delta^a$ interaction. 
All three have renormalisable interactions with the SM and $F$, and all include a neutral component which could in principle be the DM. 
However, these have already been ruled out in e.g. Ref. \cite{Cirelli:2005uq} by a combination of the relic abundance calculation and direct detection bounds. 
Adding $F$ does not change this. 
In each case, the DM mass must be $m_Z \ll m_{\rm DM} < M_F$, since otherwise it would decay rapidly, therefore the relic abundance is determined by DM annihilations into SM particles and its annihilations into $F$ particles is unimportant.\footnote{$m_{\rm DM} > M_F$ is allowed if the particle is coupled extremely feebly to the $F$, but in this case once again the DM annihilations into $F$ particles are irrelevant. } 
In each case, the spin-independent $\rm{DM}\,  N \to \rm{DM}\, N$ cross-section mediated by the $Z$-boson is also unaffected by the existence of the $F$. 
We note that the direct detection constraints are considerably weaker for a SM multiplet with $Y=0$. 
However, since the hypercharge of $F$ is larger than the hypercharge of any SM field, a renormalisable coupling involving $F$, a SM field and third field is only possible if this third field has non-zero hypercharge.

There is one DM candidates which is neutral with respect to the SM: a real scalar $\phi$.\footnote{One might also consider vector boson DM, $Z'_\mu$, however strictly a massive vector boson is not a single-field extension, since an additional field is required to generate its mass. We will briefly comment on this scenario at the end of the section. }
We now consider this scalar case.


\begin{figure}
    \centering
    
\begin{tikzpicture}
    \begin{feynhand}
      \vertex[dot] (v1) at (2.1,1.5);
      \vertex[dot] (v2) at (2.1,-1.5);
        \vertex (i1) at (0,1.5) {$F$};
        \vertex (i2) at (0,-1.5) {$\bar{F}$};
        \vertex (f1) at (4,1.5) {$\phi$};
        \vertex (f2) at (4,-1.5) {$\phi$};
      \propag[fermion] (i1) to (v1);
      \propag[fermion] (v2) to (i2);
      \propag[scalar] (f1) to (v1);
      \propag[scalar]  (f2) to (v2);
      \propag[fermion] (v1) to[edge label=$F$] (v2);
    \end{feynhand}
    \end{tikzpicture}
    \hspace{1cm} 
   \begin{tikzpicture}
    \begin{feynhand}
      \vertex (a1) at (0,0) {$\phi$};
      \vertex (b1) at (4.7,1.5) {$\gamma$};
      \vertex (b2) at (4.7,-1.5) {$\gamma$};
      \vertex [dot] (v1) at (1.4,0) {};
      \vertex [dot] (v2) at (3.2,1.5) {};
      \vertex [dot] (v3) at (3.2,-1.5) {};
      \propag [scalar] (a1) to (v1);
      \propag [fermion] (v1) to[edge label=$F$] (v2);
      \propag [fermion] (v3) to[edge label=$F$] (v1);
      \propag [fermion] (v2) to[edge label=$F$] (v3);
      \propag [photon] (v2) to (b1);
      \propag [photon] (b2) to (v3);
    \end{feynhand}
  \end{tikzpicture}
  \vspace{0.1cm}
    \caption{Production of $\phi$ from $F \bar{F}$ annihilation (left) and loop-induced decay to photons (right). }
    \label{fig:phiProd}
\end{figure}
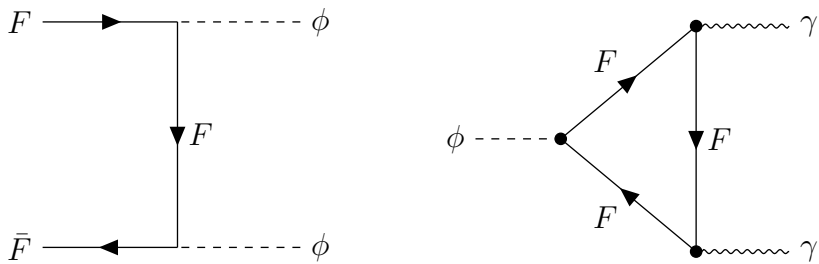

\subsubsection*{Real scalar singlet}
A real scalar singlet not only has a Yukawa interaction with the $F$, but also couples to the Higgs: 
\begin{equation}
    \mathcal{L}_\phi = \frac{1}{2}|\partial_\mu \phi|^2 - \frac{1}{2} m_\phi^2 \phi^2 - \frac{1}{6} \mu_\phi \phi^3 - \frac{1}{24} \lambda_\phi \phi^4 - \mu_{\phi H} \phi H^\dagger H - \lambda_{\phi H} \phi^2 H^\dagger H  - y_\phi \overline{F}F \phi \, .
\end{equation}
We assume that $y_\phi$ is large enough to play a role in DM phenomenology: in the limit $y_\phi \to 0$, this scenario reduces to that of Higgs-portal DM \cite{Silveira:1985rk,Burgess:2000yq} or decoupled scalar DM \cite{Arcadi:2019oxh}. 
Now we determine whether the $\phi$ can i) be produced with the correct abundance, ii) be sufficiently stable, and iii) be consistent with our understanding of structure formation.

The $\phi$ may in principle be produced via freeze-out or freeze-in, but since very small $y_\phi$ will be necessary in order to ensure the stability of $\phi$, we consider freeze-in. 
The $\phi$ is dominantly produced from the $t$-channel and $u$-channel annihilation $F \bar{F} \to \phi \phi$, see the left panel of Fig. \ref{fig:phiProd}, assuming that the $\phi$ coupling to $F$ is larger than its couplings to the Higgs. 
The thermally-averaged cross-section is computed to be
\begin{align}
    \langle \sigma(F \bar{F} \to \phi \phi) v \rangle &= \frac{y_\phi^4}{128\pi x_F^2 K_2(x_F)^2 m_F^2} \int_{4x_F^2}^\infty dr\, \frac{K_1(\sqrt{r})}{r^{3/2}} \notag \\
    &\times \left[ (r^2 + 32 x_F^4) \tanh^{-1} \sqrt{1 - \frac{4x_F^2}{r}} - 8 x_F^2 \sqrt{r^2 - 4x_F^2 r} \right] \, ,
\end{align}
where $x_F \equiv m_F/T$, neglecting $m_\phi$. 
The Boltzmann equation for freeze-in is \cite{Hall:2009bx}:
\begin{align}
    \frac{dn_\phi}{dt} + 3 H n_\phi &\simeq 2n_F^2 \langle \sigma(F \bar{F} \to \phi \phi) v \rangle \notag \\
    \Rightarrow Y_\phi(x_F) &\simeq 2 \int_0^{x_F} dx \frac{n_F^2}{s x H} \, \langle \sigma(F \bar{F} \to \phi \phi) v \rangle \, ,
    \label{eq:Yphi}
\end{align}
where $Y_\phi(x_F) \equiv n_\phi/s$ gives the yield at time $x_F$, and there is an explicit factor of 2 since each annihilation produces two $\phi$ particles. 
We take $g_* = 113.75$ including the additional relativistic degrees of freedom in the thermal plasma due to the $F$ (it equilibrates with the SM bath at temperatures $T \gtrsim M_F/20$ via gauge interactions), and since the integrand becomes exponentially suppressed for $x_F > 1$ due to the Boltzmann-suppression in $n_F$, it is a good approximation to keep $g_*$ constant. 
Furthermore, we may neglect possible number-changing $\phi \phi \leftrightarrow \phi \phi \phi \phi$ interactions, since generally the $\phi$ abundance is too small for those processes to be in equilibrium. 

Integrating Eq. \eqref{eq:Yphi} until today, $x_F \to \infty$, gives $Y_\phi(\infty) \simeq 7.5 \times 10^{10} \, y_\phi^4 (\text{TeV}/M_F)$, 
and so the relic abundance is
\begin{equation}
    \Omega_\phi h^2 \simeq 0.12 \left( \frac{y_\phi}{2.9 \times 10^{-4}} \right)^4 \frac{m_\phi}{\text{keV}} \, \frac{\text{TeV}}{M_F} \, .
\end{equation}
The seemingly large required coupling (by the standards of freeze-in), $y_\phi \sim 10^{-4}$, 
is roughly the square root of the typical freeze-in coupling, $y \sim 10^{-10}$, since the production is via $t$- and $u$-channel annihilation and therefore involves two powers of the coupling in the amplitude, as opposed to a decay or contact-interaction annihilation, both of which involve only one power of the coupling. 
For larger values of $y_\phi$, the correct relic abundance may be achieved via freeze-out.

Now we turn to the possible decays of $\phi$. 
If $m_\phi > 2M_F$, it will decay with rate $\Gamma(\phi \to F \bar{F}) = y_\phi^2 m_\phi (1- 4m_F^2/m_\phi^2)^{3/2}/(8\pi)$. 
In order for the $\phi$ to be stable for longer than the age of the Universe, $y_\phi \lesssim 10^{-22}$ is required given $m_\phi \gtrsim$ TeV. 
This is clearly inconsistent with the value of $y_\phi$ required for freeze-in of the DM. 
For $m_\phi < 2M_F$, the most pertinent decay is the loop-induced process, $\phi \to \gamma \gamma$, see the right panel of Fig. \ref{fig:phiProd}, which is always kinematically allowed. 
Its partial width is
\begin{equation}
    \Gamma(\phi \to \gamma \gamma) = \frac{25e^4 y_\phi^2 m_\phi^3}{18\pi (16\pi^2)^2 M_F^2} \, ,
\end{equation}
summing over the contributions of the singly- and doubly-charged components of $F$ in the loop. 
This corresponds to a lifetime of 
\begin{align}
    \tau_{\phi \to \gamma \gamma} = 5.2 \times 10^{13} \left( \frac{2.9 \times 10^{-4}}{y_\phi} \right)^2 \left( \frac{\text{keV}}{m_\phi} \right)^3 \left( \frac{M_F}{\text{TeV}}\right)^2 \, \text{s} \, .
\end{align}
Lyman-$\alpha$ observations, which probe small-scale structure, forbid frozen-in dark matter lighter than around 15 keV \cite{Decant:2021mhj}. 
On the other hand, scalar DM decaying into a pair of photons has been constrained in Ref. \cite{Essig:2013goa} to have a lifetime of more than $\sim 10^{26}$s, which is only allowed by an extremely large fermion mass, $M_F \gtrsim  (m_\phi/\text{keV})^{3/2} \text{ EeV} $: for $m_\phi \gtrsim 15$ keV and $y_F \lesssim 4\pi$ this implies $\epsilon_{F\alpha} \lesssim 3 \times 10^{-8}$. 
Such a small value of $\epsilon_F$ will not be probed even by future $\mu \to e$ searches, as can be seen from Fig. \ref{fig:emu}. 
Moreover this scenario also requires very small $\phi-H$ couplings, in particular $\lambda_{\phi H} < m_\phi^2/v^2$. 
However, the possibility of detecting $\phi \to \gamma \gamma$, or indeed other $\phi$ decays such as $\phi \to \nu \overline{\nu}$ and $\phi \to e^+ e^-$, which should have similar partial widths, is a notable feature.

We note now that the vector DM case, which we did not consider, has similar phenomenology: the $Z'$ must be frozen-in, and requiring that the rates of the decays $Z' \to \nu \overline{\nu}$ and $Z' \to e^+ e^-$ are slow enough to avoid existing bounds implies that $\epsilon_{F\alpha} \ll 1$.

Thus, the single-field option is possible but is undoubtedly contrived: it requires several parameters to be very tiny without any justification. 
In that sense, it is no more than a proof of principle of the most minimal $F$-portal DM. 
One may very well prefer a DM model without any small couplings. 
In order to achieve this, we turn to a scenario with the $F$ plus two additional fields.

\subsection{Two field extensions}
%
There are perhaps many possible two-field extensions which give a satisfactory DM candidate. 
To give a concrete example, we introduce a scalar, $\Phi \sim (1,2)_{-3/2}$, and a vector-like singlet fermion, $\chi \sim (1,1)_0$, which is the DM. 
The Lagrangian is
\begin{equation}
	\mathcal{L} = \mathcal{L}_{\text{SM}+F} + \overline{\chi} (i \slashed{\partial} - m_\chi) \chi + |D_\mu \Phi|^2 - y_\Phi \overline{\chi} \Phi^\dagger F - y_\chi \overline{\chi} \tilde{H}^\dagger l_L - V(H,\Phi) + h.c. \, ,
\end{equation}
where $V(H,\Phi)$ is the scalar potential. 
The Yukawa coupling $y_\chi$, well-known from the type-I seesaw Lagrangian, must be very small so as not to give too large a contribution to neutrino masses. 
We therefore set $y_\chi \to 0$: the simplest way to forbid this term is via a $\mathbb{Z}_2$ symmetry under which $\Phi \to - \Phi$ and $\chi \to -\chi$ while all other fields are uncharged. 
This moreover stabilises the DM so long as $m_\chi < m_\Phi + M_F$. 
The $\mathbb{Z}_2$ could be a subgroup of a larger gauge symmetry. 
The interactions of the $\Phi$ are much less troublesome: assuming that $m_\Phi \gtrsim M_F$, it has a negligible impact on phenomenology.

\begin{figure}[t!]
    \centering
    \includegraphics[width=0.65\textwidth]{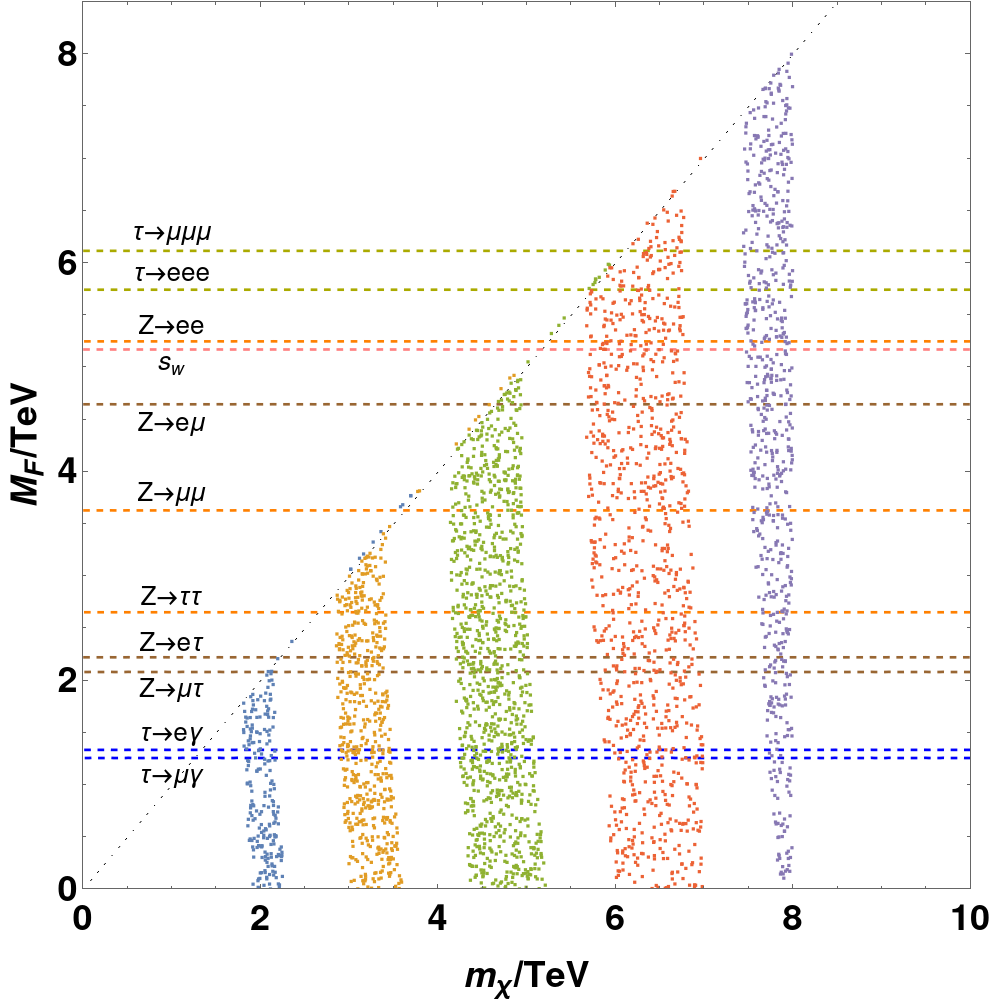}
    \caption{Scan of points which produce a DM abundance, $\Omega_\chi h^2 \in [0.1, 0.14]$. 
    The blue, orange, green, red and purple correspond to $m_\Phi/y_\Phi = 4,5,6,7,8$ TeV, respectively. In each case, we enforce that $M_F, m_\chi < m_\Phi/y_\Phi$. 
    The horizontal lower bounds on $M_F$ come from observables considered in Section \ref{sec:pheno}, assuming $y_{Fe} = y_{F\mu} = y_{F\tau} = 1$, and follow the colour scheme of Fig. \ref{fig:emu}. These weaken linearly as $y_F$ decreases. } 
    \label{fig:DM}
\end{figure}

Now we consider DM production. 
From hereon we will assume that $m_\Phi > M_F$, but little would change if $M_F > m_\Phi$. 
Both the $F$ and $\Phi$ will thermalise with the SM particles due to their gauge interactions. 
If $y_\Phi$ is small, the $\chi$ does not thermalise but is instead frozen-in predominantly via the decays $\Phi \to \overline{\chi} F$ and $\Phi^\dagger \to \chi \overline{F}$. 
This is an acceptable production mechanism, however in this section we aim to avoid very small couplings, so we consider $y_\Phi$ to be roughly $\mathcal{O}(1)$, in which case $\chi$ will thermalise with the bath of SM particles, $F$ and $\Phi$. 
Later, its abundance is frozen out via $\Phi$-mediated $\chi \bar{\chi} \to F \bar{F}$ annihilations. 
If $m_\chi > M_F$, the relic density of $\chi$ is determined by the standard freeze-out calculation. 
On the other hand, if $m_\chi < M_F$, it is a scenario of forbidden DM, wherein the DM annihilates into heavier particles \cite{Griest:1990kh}. 
The relevant annihilation rate in the limit $m_\Phi \gg m_\chi, M_F$ is
\begin{align}
    \sigma(\chi \overline{\chi} \to F \overline{F}) &= \frac{y_\Phi^4}{48\pi s m_\Phi^4} \sqrt{\frac{s-4M_F^2}{s-4m_\chi^2}} \left[ s^2 + s (6M_F m_\chi - M_F^2 - m_\chi^2) +16 M_F^2 m_\chi^2 \right]
    \label{eq:DMcross}
\end{align}
In Fig. \ref{fig:DM}, we plot the result of a scan of points in the $M_F$ vs. $m_\chi$ plane which give $\Omega_\chi h^2 \in [0.1, 0.14]$ for different values of $m_\Phi/y_\phi$ up to 8 TeV, which is around the largest value for which the correct abundance can be obtained. 
We have imposed that $m_\Phi/y_\Phi > m_\chi, M_F$, so that for large $y_\Phi$ we recover the assumption on the hierarchy of masses made in deriving Eq. \eqref{eq:DMcross}. 
Note that almost all points are in the standard freeze-out case, $m_\chi > M_F$, with only a few around the line $m_\chi \simeq M_F$ and none with $M_F$ clearly larger than $m_\chi$. 
This is a result of the exponential sensitivity to the degree of mass-splitting, $(M_F-m_\chi)/m_\chi$, which is a characteristic feature of forbidden DM. 
The horizontal lines correspond to bounds on $M_F$ from different observables assuming $y_{Fe} = y_{F\mu} = y_{F\tau} = 1$: they weaken linearly as $y_F$ decreases. 
Constraints from $\mu \to e$ transitions are even more stringent: these are compatible with the DM scenario only if either the flavour structure of the new physics forbids $\mu \to e$, or if at least one of $y_{Fe}$ and $y_{F\mu}$ is $\ll 1$. 
Since $M_F$ must be at least a few hundred GeV, there are good prospects for detecting or ruling out this DM scenario.

In summary, the $F$-portal provides a diversity of possible DM candidates. 
We have demonstrated the possibility of light, bosonic DM produced via freeze-in of $F \overline{F}$ annihilations, as well as heavy fermionic DM which freezes-out via $\chi \overline{\chi} \to F \overline{F}$. 
In the former case, the DM is the only new field and may be detected via decays to photons and possibly also to SM fermions, while requiring that $M_F$ is EeV-scale or heavier. 
In the latter case, where the DM is accompanied by a second new field, it is more difficult to detect even indirectly since it is stable. 
However, as shown in Fig. \ref{fig:DM}, this scenario requires a light $M_F$, which may be detected in the future either at colliders or indirectly via the measurements discussed in Section \ref{sec:pheno}. 
While more complicated DM models involving the $F$-portal can no doubt be constructed, the cases described above are notable for their minimality and accessible phenomenology.

\section{Conclusion}
\label{sec:conc}
This paper has addressed a simple question: what are the consequences of extending the SM by the weak-doublet fermion, $F$, via the $(\overline{F} \tilde{H} e_R + h.c.)$ interaction? 
This field has previously received scant attention in the literature.

The study was broken down into two parts. 
Firstly, in Sections \ref{sec:model} and \ref{sec:pheno} we found the dim-6 EFT generated by integrating out the $F$ and used this to constrain the model from a variety of leptonic and EW observables. 
The rates of different processes are highly correlated due to the small number of free parameters in the model. 
As shown in Figs. \ref{fig:emu}-\ref{fig:mutau}, the most constraining current and expected future observables are $\mu \to e$ conversion in nuclei and $\mu \to 3e$, while flavour-violation in the $\tau \to e$ and $\tau \to \mu$ sectors is bounded at approximately the same level as various flavour-conserving processes. 
In contrast with many other models, the SM$+F$ affects flavour-conserving $Z$-boson decays and the weak-mixing angle, but not the $W$-boson mass. 
Secondly, in Section \ref{sec:DM}, we showed how extending the model by a single field gave a keV scalar DM candidate with possibly detectable decays into photons, while extending it by two fields lead to a fermionic DM candidate which necessarily implied that $F$ could not be heavier than a few TeV, and thus perhaps detectable in the next generation of experiments.

This analysis opens up several possible future directions. 
Sticking firstly to the fermion $F$, a natural question to ask is how to embed this field within a larger model of new physics. 
This was attempted in a minimal way in Section \ref{sec:DM} with regards to dark matter. One could also attempt to build, for instance, neutrino mass models with the $F$ or explain the matter-antimatter asymmetry via a scenario of `$F$-genesis'. 
Such model-building efforts would of course be most motivated if there were future hints for the $F$ fermion: as noted above, the first signal is likely to be in $\mu \to e$ conversion, flavour-conserving $Z$ decays or the weak-mixing angle.

More generally, this study aims to encourage further investigation into simple but relatively unexplored ideas which may have suffered from some theoretical prejudice. 
One can work systematically, starting with single-field extensions and building up. 
This differs from a general EFT analysis, which has the same goal of performing an agnostic survey of the parameter space of possible new physics. 
Firstly, the new physics does not necessarily have to be well above the EW scale (or some other cutoff scale); secondly, the parameter space is far more tractable than is often the case in EFTs, in the sense that a given model typically has many fewer parameters than a general EFT.

\vspace{1cm}
\noindent \textbf{Acknowledgements}\\
I thank Michele Frigerio for his very helpful comments on the manuscript. 
I also thank Quentin Decant for useful conversations on \cite{Decant:2021mhj} and Wolfgang Altmannshofer for clarifying an issue in \cite{Altmannshofer:2013zba}.
This project has received support from the IISN convention 4.4503.15. 


\bibliography{frogsbib.bib}


\end{document}